\documentstyle[12pt]{article}
\newcommand{\bmalpha}{\mbox{{\boldmath $\alpha$}}}

\begin{document}

\begin{flushright}
NIIG-DP-98-3\\
December, 1998\\
hep-lat-9812002
\end{flushright}

\begin{center}
{\Large Ginsparg-Wilson Relation and Lattice Supersymmetry}\\
\vskip .75in

{\large Hiroto So\footnote{so@muse.hep.sc.niigata-u.ac.jp} and Naoya Ukita\footnote{ukita@muse.hep.sc.niigata-u.ac.jp}}\\
\vskip .2in
{\it Department of Physics},\\
{\it Niigata University},\\
{\it Ikarashi 2-8050,
 Niigata 950-2181, Japan.}\\

\vskip .5in
\end{center}
\begin{abstract}
\baselineskip 16pt
The Ginsparg-Wilson(G-W) relation  is extended for 
  supersymmetric free theories on a lattice.  
 Exact  lattice supersymmetry(SUSY) can be defined without 
 any  ambiguities in difference operators.  
 The  lattice action constructed by a block-spin transformation 
 is invariant under the symmetry.  
  $U(1)_R$ symmetry on the lattice is also realized 
as one of exact symmetries.
 For an application, the  extended G-W relation is given 
 for a two-dimensional model with chiral-multiplets.
  It is argued that the relation  may be  generalized for interacting cases. 
\end{abstract}

\setcounter{footnote}{0}

\newpage

{\bf 1.~}{\it Introduction} \     
Recently, there have been much progress in understanding 
chiral symmetry on a  lattice.
  L{\"u}scher  has given a lattice form of a 
  chiral transformation, 
  which differs from the one in the continuum theory
 by a term proportional to lattice spacing\cite{lu}.
 This result may contain an important step 
 to escape the no-go theorem\cite{nogo}.
 The lattice fermion action is invariant  
  under the lattice chiral symmetry 
 if   the Dirac operator  satisfies   
  the Ginsparg-Wilson(G-W)  relation\cite{gw}.
  The relation  should be recognized as 
 a remnant of the continuum chiral symmetry 
 and a kind of Ward-Takahashi identity. 
  A particular  solution 
has been given by Neuberger\cite{neu}.

In this paper, we aim at the construction 
 of an exact lattice supersymmetry(SUSY) 
 consistent with   the SUSY-extension of the  G-W relation. 
  The first step to get the  exact SUSY 
 in the spirit of  L{\"u}scher  is 
    to restate a continuum SUSY as a  
 na{\"\i}ve lattice SUSY        without 
 any ambiguities in  difference operators. 
 To this end, we must introduce 
 a block-spin function which transforms field variables in 
 the continuous space  into   dynamical ones on the lattice.
 The second step is to define an exact  SUSY transformation  from 
 the na{\"\i}ve lattice version, where we use 
 our  SUSY-extension of the G-W relation.  
 Our final step is  to determine an arbitrary parameter 
  in  the definition of  the exact  SUSY transformation.
 It should be consistent with the lattice algebra.
Furthermore,   we find a $U(1)_R$ charge on the lattice in the spirit of  
L{\"u}scher. There appears no arbitrary parameter in the charge.
Finally, we find exact  SUSY and $U(1)_R$ invariance  of lattice 
theories without doubling problems. 

 In order to see how our approach works, we construct 
  a supersymmetric free  theory  in two dimensions. 
  In addition to  fermionic and bosonic G-W relations 
used by Kikukawa-Aoyama\cite{ak},  we can derive the more relations
 for the lattice action. Once one of kinetic terms for component fields 
 is found, we can construct a SUSY-invariant total action.

 Our approach can be generalized  for interacting cases,
owing to the SUSY-extension of the  G-W relation.    
 Unlike previous works on lattice SUSY[5-7], we need not  
 to introduce  the definition 
 of difference operators by hand.
 Therefore, the problem of Leibniz rule may   be overcome.

{\bf 2.~}{\it Derivation of SUSY Ginsparg-Wilson relation } \ 
 The original G-W relation was derived as an identity 
 for  the Gaussian type  effective action.
 The relation may play an important role in characterizing a
lattice analog of the chiral symmetry which should be realized in the
vicinity of the continuum limit.   

A way out of the no-go theorem is to introduce two  chiral 
 matter fields.  
  A SUSY transformations for  
two chiral-multiplets 
 $\Phi_j = 
 (\phi_j, \psi_j, F_j)^{\rm T}$ $j=1,2$   in the continuum theory are 
 defined as

\begin{equation}
\delta_{\epsilon} \Phi_j =  Q(\epsilon , \bar{\epsilon})\Phi_j ,
\end{equation}

\begin{equation}
\delta_{\epsilon} \bar{\Phi}_j = \bar{\Phi}_j  \bar{Q}(\epsilon ,
\bar{\epsilon}) ,
\end{equation}

\noindent 
where T represents a transpose operation and 
 the space-time is assumed to be Euclidean.
 In the rest of the section, we 
suppress the index $j$ for  chiral-multiplets.
 Starting with a continuum theory, we define its regularized theory on a 
cubic lattice by performing a block-spin transformation. 
 A lattice point is expressed by an integer vector  
\{$n_{\mu}a$\}, where $a$ is lattice constant. We take $a = 1$
 for simplicity. 
 The block-spin transformation from $\Phi (x)$ to $\Phi_n$ is given by

\begin{equation}
\Phi_n  \sim \displaystyle{ \int} {\rm d}x f_n(x) \Phi (x) ~~ \equiv 
~~\langle f_n, \Phi \rangle ,
\end{equation}

\begin{equation}
\bar{\Phi}_n  \sim \displaystyle{ \int} {\rm d}x f_n(x) \bar{\Phi} (x)
 ~~ \equiv 
~~\langle f_n, \bar{\Phi} \rangle ,
\end{equation}

\noindent 
where $f_n(x) =  f(x-n)$ is  a block-spin function 
 with finite support around
 $x_{\mu} = n_{\mu}$.
  $\langle , \rangle$ implies the usual inner product in a function 
space. 

Following  to  Ginsparg
and Wilson\cite{gw},
we may define a Gaussian effective action $A_{\rm eff}$ by using a
 SUSY-invariant massless  
action $A_c$ in the continuum theory\footnote{Since we have two chiral 
multiplets,   it is possible to construct a Dirac mass term.}:

$$
\displaystyle{\exp (-A_{\rm eff}[\Psi_n, \bar{\Psi}_n])} 
$$
\begin{equation}
  =  \int {\cal D}\Phi (x)  {\cal D}\bar{\Phi} (x)  \exp ( - \sum_{n,m}(\bar{\Psi}_n - \bar{\Phi}_n)
\bmalpha_{n,m}
(\Psi_m - \Phi_m)    - A_c [\Phi, \bar{\Phi}]) . 
\end{equation}

\noindent
Here $\alpha_{n,m}$ is a matrix acting on the  multiplet $\Psi_n$,

\begin{eqnarray}
\bmalpha_{n,m} = \alpha \  \delta_{n,m} \left(\begin{array}{rrr}
0 & 0 & 1 \\
0 & V & 0 \\
 1 & 0 & 0 
\end{array}
\right)  ,
\end{eqnarray}

\noindent
where $V$ is some anti-symmetric matrix determined by a mass-term
 in a SUSY-invariant Lagrangian and 
   $\alpha$ is proportional to $O(a^{-1})$.

We may define na{\"\i}ve lattice SUSY by restating the continuum SUSY 
  as follows:

\begin{equation}
\begin{array}{lll}
\delta^N_{\epsilon} \Phi_n & = & \displaystyle{\int }  f_n (x) \delta \Phi (x)
{\rm d}x \\
 & & \\
 & = & Q_L(\overrightarrow{\bigtriangledown}) \Phi_n  ,
\end{array}
\end{equation}

\begin{equation}
\begin{array}{lll}
\delta^N_{\epsilon} \bar{\Phi}_n & = & \displaystyle{\int }  f_n (x) \delta \bar{\Phi} (x)
{\rm d}x \\
 & & \\
 & = &  \bar{\Phi}_n \bar{Q}_L(\overleftarrow{\bigtriangledown})  .
\end{array}
\end{equation}

\noindent
A  derivative operator in the continuum SUSY 
 is replaced by a difference operator in the lattice SUSY
 using the relation $\partial_{\mu} f_n(x) = - 
\overrightarrow{\bigtriangledown} _{\mu}f_n(x)$. 
 Although the explicit form of the  difference operator depends on the 
 block-spin function, 
  it is possible to choose  a reasonable function  for  
 the realization of  lattice SUSY. 
 As a preparation for the exact lattice SUSY, we would just look 
 at some properties of the  na{\"\i}ve SUSY. 

Under this na{\"\i}ve transformation,  
 our effective action changes by

$$
\exp (-A_{\rm eff}[\Psi ', \bar{\Psi} ']) 
$$
\begin{eqnarray}
  = & \displaystyle{\int} {\cal D}\Phi (x) {\cal D}\bar{\Phi} (x)
 \exp ( - (\bar{\Psi}' - \bar{\Phi})\bmalpha
(\Psi' - \Phi)    - A_c [\Phi , \bar{\Phi}])  \nonumber  \\[5mm]
 =  &  \displaystyle{\int} {\cal D}\Phi (x) {\cal D} \bar{\Phi} (x)
 \exp ( - (\bar{\Psi} - \bar{\Phi}') e^{\bar{Q}_L}\bmalpha
 e^{Q_L}(\Psi - \Phi')    - A_c [\Phi ', \bar{\Phi} '])  ,
\end{eqnarray}

\noindent
where the lattice site and the spinor indices are both omitted. It is
assumed that the  $A_c$ 
 is invariant under  SUSY transformation in the continuum theory,

\begin{equation}
A_c[\Phi , \bar{\Phi}] = A_c[\Phi ', \bar{\Phi} '] .
\end{equation}

\noindent
Although the path-integral measure is  na{\"\i}vely unchanged

\begin{equation}
{\cal D}\Phi'{\cal D}\bar{\Phi}' = {\cal D}\Phi {\cal D}\bar{\Phi} ,
\end{equation}

\noindent
we take account of the contribution of the Jacobian factor which is
needed to consider a possible effect of the anomaly\cite{fuji}.

Let us derive a SUSY extension of the G-W relation for a free theory
 described by

\begin{equation}
A_{\rm eff}[\Psi , \bar{\Psi}] = \sum_{n,m} 
\bar{\Psi}_n S_{(n,m)} \Psi_m  .  
\end{equation}

\noindent
Under the  na{\"\i}ve SUSY , it  transforms as

$$
\exp (-A_{\rm eff}[\Psi , \bar{\Psi}])(1 -  (\bar{\Psi} (S Q_L +
\bar{Q}_L S)\Psi 
) 
$$
\begin{eqnarray}
= & (1 + \delta J - {\rm str}~\bmalpha^{-1}
( \bmalpha Q_L + \bar{Q}_L\bmalpha )
\bmalpha^{-1}S
+ {\rm str}~\bmalpha^{-1}
( \bmalpha Q_L + \bar{Q}_L\bmalpha )
 &  \nonumber  \\ [4mm]
  & -
\bar{\Psi}S 
  \bmalpha^{-1}
( \bmalpha Q_L + \bar{Q}_L\bmalpha )\bmalpha^{-1} 
S\Psi   ) \exp (-A_{\rm eff}[\Psi , \bar{\Psi}])  ,
 &                          
\end{eqnarray}

\noindent
where 
 $\delta J$ comes from 
 a Jacobian factor.
 So, we can get following two relations: 

\begin{equation}
  \delta J = {\rm str}~\bmalpha^{-1}( \bmalpha Q_L(\overrightarrow{\bigtriangledown}) +
\bar{Q}_L(\overleftarrow{\bigtriangledown})\bmalpha )
 \bmalpha^{-1}S - {\rm str}~\bmalpha^{-1}
( \bmalpha Q_L(\overrightarrow{\bigtriangledown}) +
\bar{Q}_L(\overleftarrow{\bigtriangledown})\bmalpha ) ,
\end{equation}

\noindent 
and

\begin{equation}
 \bar{\Psi} (S Q_L(\overrightarrow{\bigtriangledown}) + \bar{Q}_L(\overleftarrow{\bigtriangledown}) S)\Psi  = 
\bar{\Psi}S 
   \bmalpha^{-1}
 ( \bmalpha Q_L(\overrightarrow{\bigtriangledown}) + \bar{Q}_L(\overleftarrow{\bigtriangledown})\bmalpha ) 
\bmalpha^{-1} 
S\Psi  . \label{gw2}
\end{equation}

\noindent
These are SUSY extended G-W relations. Note that the  right
hand sides of these relations vanish if the difference operator($\overrightarrow{\bigtriangledown}$) is
anti-symmetric in the  matrix notation.

For an exact lattice SUSY transformation, the above relations suggest us
to define

\begin{equation}
q \equiv Q_L(\overrightarrow{\bigtriangledown}) -
Q_L(\overrightarrow{\bigtriangledown}) 
 \bmalpha^{-1}S 
\end{equation}
\begin{equation}
\bar{q} \equiv  \bar{Q}_L (\overleftarrow{\bigtriangledown})
 -  S \bmalpha^{-1} \bar{Q}_L (
\overleftarrow{\bigtriangledown})
\end{equation}

\noindent
under which we can show the invariance of our effective action:

\begin{equation}
 \delta A_{\rm eff} = \bar{\Psi} (S q  + \bar{q} S)\Psi  = 0 .
\end{equation}

\noindent
This is a SUSY extension of  L{\"u}scher's symmetry\cite{lu}.
 Similar to chiral symmetry, SUSY is also modified by $O(a)$ because of 
$\bmalpha = O(a^{-1})$.
  We must note that there may exist an arbitrary parameter $c$ in 
 defining lattice SUSY:

\begin{equation}
q_c \equiv q - c
 Q_L ( \overrightarrow{\bigtriangledown}_{\rm s}) 
 \bmalpha^{-1}S ,
\end{equation}
\begin{equation}
\bar{q}_c \equiv  \bar{q}
 - c  S \bmalpha^{-1} \bar{Q}_L ( 
\overleftarrow{\bigtriangledown}_{\rm s})  ,
\end{equation}
\noindent 
where the symmetric difference operator 
$\overrightarrow{\bigtriangledown}_{\rm s} \equiv 
{\displaystyle \frac{\overrightarrow{\bigtriangledown}
-\overrightarrow{{\bigtriangledown}}{} ^{\rm T}}{2}}$.
 The parameter  should be determined by
 the closure property of the algebra.
 Since this transformation is not unitary, we  get
 the Jacobian factor $1 - \delta J$ under  the transformation Eqs.(19)
and (20); $\delta J$ 
 is expressed as Eq.(14) and  is easily shown to be 
 independent of the parameter $c$.

 For $U(1)_R$ charge  $Q_R$, we  find the G-W relations 
 similar to the SUSY case: 

\begin{equation}
  \delta J_R = {\rm str}~\bmalpha^{-1}( \bmalpha Q_R +
\bar{Q}_R\bmalpha )
 \bmalpha^{-1}S - {\rm str}~\bmalpha^{-1}
( \bmalpha Q_R +   \bar{Q}_R \bmalpha ) ,
\end{equation}

\noindent 
and

\begin{equation}
 \bar{\Psi} (S Q_R + \bar{Q}_R S)\Psi  = 
\bar{\Psi}S 
   \bmalpha^{-1}
 ( \bmalpha Q_R + \bar{Q}_R\bmalpha ) 
\bmalpha^{-1} 
S\Psi  . 
\end{equation}

\noindent
The conserved `$U(1)_R$' charge \cite{ak}, 

\begin{equation}
q_R \equiv Q_R ( 1 - 
 \bmalpha^{-1}S ) ,
\end{equation}

\noindent
can be also found and has no arbitrary parameter  unlike the  SUSY case.

{\bf 3.~}{\it An example: 2-Dimensional chiral-multiplets} \  
 We consider two  chiral-multiplets  $\Phi_j$, $j=1,2$, consisted 
of real scalars $\phi_j$,  auxiliary fields $F_j$ and complex Weyl spinors $\chi_j$. 
 These are arranged in 
  a complex multiplet $\Phi = (\phi_1 + i\phi_2, \chi_1 + i\chi_2,
 \chi^*_1 + i \chi^*_2, F_1 + iF_2)^{\rm T} \equiv
 (\phi, \chi, \bar{\chi}, F)^{\rm T}$ and its conjugate $\bar{\Phi} =
 (\phi_1 -i\phi_2, \chi_1 - i\chi_2,
\chi^*_1 - i \chi^*_2, F_1 - iF_2) ) \equiv (\phi^*, \bar{\chi}^{\dagger}, 
\chi^{\dagger}, F^*) $.
 We define N=1 SUSY transformation:

\begin{equation}
 \left\{ \begin{array}{l}
\delta_\epsilon \phi = i (\epsilon ^* \chi  + \epsilon  \bar{\chi})\\
\delta_\epsilon \chi = - 2   \epsilon  ^* \partial_z \phi + i \epsilon F \\
\delta_\epsilon \bar{\chi} =  - 2   \epsilon  \partial_{\bar{z}} \phi  -
 i\epsilon ^* F \\
\delta_\epsilon F =  - 2  \epsilon  \partial_{\bar{z}} \chi + 2  \epsilon ^* 
\partial_z \bar{\chi} .
\end{array} \right. 
\end{equation}

\noindent
We consider a SUSY-invariant massless  Lagrangian:

\begin{equation}
{\cal L} =  2 \partial_{\bar{z}}\phi^* \partial_z\phi + 
i (\chi ^{\dagger} \partial_z \bar{\chi} + 
\bar{\chi}^{\dagger} \partial_{\bar{z}}\chi) -  {\displaystyle
\frac{1}{2}}F^* F .
\end{equation}

\noindent 
Then, we obtain the matrix $V$ in  Eq.(6),

$$
V = \left(\begin{array}{cl}
  0 & -1 \\
 1 & 0
 \end{array}\right) ,
$$

\noindent
from  a mass-term in a SUSY-invariant Lagrangian,

\begin{equation}
{\cal{L}}_m  =  - \frac{m}{2} (F^*\phi + F \phi ^* + \chi^{\dagger} \chi - \bar{\chi}^{\dagger} \bar{\chi}) . 
\end{equation}

The na{\"\i}ve lattice SUSY takes of the matrix form:

\begin{equation}
Q_L = \left(\begin{array}{cccc}
 0 & i \epsilon ^* & i \epsilon & 0 \\
 - 2 \epsilon ^*   \overrightarrow{\bigtriangledown}_z & 0 & 0 &  i \epsilon \\
 - 2 \epsilon    \overrightarrow{\bigtriangledown}_{\bar{z}} & 0 & 0 & - i \epsilon ^*  \\
  0 &  - 2 \epsilon    \overrightarrow{\bigtriangledown}_{\bar{z}} & + 2 \epsilon ^*  \overrightarrow{\bigtriangledown}_z & 0 
 \end{array}
 \right) ,
\end{equation}

\noindent
and 

\begin{equation}
\bar{Q}_L = \left(\begin{array}{cccc}
 0 & - 2 \epsilon ^* \overleftarrow{\bigtriangledown}_z &  - 2 \epsilon 
 \overleftarrow{\bigtriangledown}_{\bar{z}} & 0 \\
 - i \epsilon ^* & 0 & 0 & 2 \epsilon 
 \overleftarrow{\bigtriangledown}_{\bar{z}} \\
 - i \epsilon & 0 & 0 &  - 2 \epsilon ^*
\overleftarrow{\bigtriangledown}_z \\
 0 &  i \epsilon  & - i \epsilon ^* & 0 
 \end{array}
 \right)   .
\end{equation}

\noindent
It follows that

\begin{equation}
\bmalpha Q_L + \bar{Q}_L\bmalpha  = \alpha \left(\begin{array}{cccc}
0 & -2 \epsilon (TD)_{\bar{z}} & 2 \epsilon^* (TD)_{z} & 0 \\
 2 \epsilon (TD)_{\bar{z}} & 0 & 0 & 0 \\
-2 \epsilon^* (TD)_{z} & 0 & 0 & 0 \\
0 & 0 & 0 & 0 
 \end{array}
 \right) ,
\end{equation}

\noindent
where $ (TD)_z$ denotes  a total derivative
$\overrightarrow{\bigtriangledown}_z + 
\overleftarrow{\bigtriangledown}_z$ .

We obtain in our approach not only the original   G-W relation
  but also the relation among  kinetic terms for  
fermion, boson and  auxiliary fields:

\begin{equation}
 |S_{\bar{\chi}^{\dagger} ,\chi}|^2 + |S_{\chi^{\dagger} ,\chi}|^2
 = - \alpha  S_{\chi^{\dagger},\chi} ,
\end{equation}

\begin{equation}
S_{F^*,F} S_{\phi^*,\phi} + |S_{F^*,\phi}|^2
 = \alpha  S_{F^*,\phi} ,
\end{equation}

\begin{equation}
2  S_{F^*,F}(TD)_{\bar{z}}S_{\chi^{\dagger},\chi}
 = \alpha (-2 S_{F^*,F} \overrightarrow{\bigtriangledown_{\bar{z}}} +
iS_{\bar{\chi}^{\dagger} ,\chi})  ,
\end{equation}
      
\begin{equation}
2  S_{\phi^*,F}(TD)_z S_{\bar{\chi}^{\dagger},\chi}
 = i \alpha ( S_{\phi^*, \phi} + 2i
\overleftarrow{\bigtriangledown_z}S_{\bar{\chi}^{\dagger} ,\chi}) .
\end{equation}

\noindent
Once we find the fermion kinetic term, $S_{\bar{\chi}^{\dagger} ,\chi}$,
  it is easy to express the total  action  explicitly
from these relations.

{\bf 4.~}{\it Conclusions} \  
  ~~~ In this letter, 
  we have extended 
the Ginsparg-Wilson relation for a supersymmetric case and 
 applied it for free theories\footnote{Bietenholz 
also studied  SUSY-extended G-W relations 
by similar approach \cite{wb}, but  his  expression for  G-W relations
 is quite different from ours.}.
  The exact  lattice SUSY  and $U(1)_R$ symmetry  are found. 
 As an example, SUSY-extended G-W relations  of 
2-dimensional two chiral-multiplets 
  are derived.  Similarly, for a Majorana fermion, 
one can describe a SUSY 
theory  by  single non-chiral matter field escaping from the no-go
theorem.

 Our block-spin construction uniquely fixes 
the definition of difference operators 
 owing to our block-spin construction.
 If one uses 
the even block-spin function which leads us to  
$\bmalpha Q_L + \bar{Q}_L\bmalpha = 0$,  
 our lattice SUSY algebra  is exactly closed.

 For interacting cases, we can also derive the G-W relation.  
 If we use even block-spin functions, the  na{\"\i}ve SUSY is an exact 
 symmetry for the block-spin effective action and 
 we may  overcome  the problem of  Leibniz rule. 
 Further   investigation to  the G-W relation is necessary to show 
 these properties in concrete terms.
  Finally, this  G-W relation is generally important for understanding 
 the continuum limit with the supersymmetry.    
 The detailed  analysis shall be reported  in a forthcoming  paper\cite{ksu}.

 We are grateful to Y. Igarashi for reading our manuscript 
 carefully and invaluable comments.
 We also thank 
 K. Itoh and H. Nakano for their encouragement.


\begin{thebibliography}{99}
\bibitem{lu} M.~L{\"u}scher, Phys.\ Lett. {\bf B428} (1998) 342.
\bibitem{nogo}  H.~B.~Nielsen and M.~Ninomiya, Nucl.\ Phys.\ {\bf
 B185} (1981) 20;
  {\bf B195} (1982) 541;  {\bf B193} (1981) 173.
\bibitem{gw} P.~Ginsparg and K.~Wilson, Phys.\ Rev.\ {\bf D25} (1982) 2649.
\bibitem{neu} H.~Neuberger, Phys.\ Lett.\ {\bf B417} (1998) 141;  
 {\bf B427} (1998) 353.
\bibitem{ak} T.~Aoyama and Y.~Kikukawa, Phys.\ Rev.\ {\bf D59} (1999) 054507.
\bibitem{ss} N.~Sakai and M.~Sakamoto, Nucl.\ Phys.\ {\bf B229} (1983) 173.
\bibitem{sn} S.~Nojiri, Prog.\ Theor.\ Phys.\ {\bf 74} (1985) 819.
\bibitem{fuji} K.~Fujikawa, Phys.\ Rev.\ {\bf D21} (1980) 2848;
  {\bf D22} (1980) 1499.  
\bibitem{wb} W.~Bietenholz, Mod.\ Phys.\ Lett. {\bf A14} (1999) 51.
\bibitem{ksu} M.~Koseki, H.~So and N.~Ukita, in preparation.
\end{thebibliography}
\end{document}